# An energy-efficient scheduling algorithm for shared facility supercomputer centers

E.A. Kiselev[1], P.N. Telegin[1], B.M. Shabanov[1]

[1]Joint Supercomputer Center of the Russian Academy of Sciences – Branch of Federal State Institution Scientific Research Institute for System Analysis of the Russian Academy of Sciences, Leninsky prospect, 32a, Moscow, 119334, Russian Federation

**Abstract**

The evolution of high-performance computing is associated with the growth of energy consumption. Performance of cluster computes (CC) is increased via rise in performance and the number of used processors, GPUs and coprocessors. Increment in number of computing elements results significant growth of energy consumption. Power grids limits for supercomputer centers (SCC) are driving the transition to more energy efficient solutions. Often upgrade of computing resources is done step-by-step, i.e. parts of older supercomputers are removed from service and replaced with newer ones. A single SCC at any time can operate several computing systems with different performance and power consumption. That's why the problem of scheduling parallel programs execution on SCC resources to optimize energy consumption and minimize the increase in execution time (energy efficient scheduling) is important. The goal of presented work was development of a new energy efficient algorithm of scheduling computing resources at SCC. To reach the goal the authors analyzed methods of scheduling computing resources in shared facility, including energy consumption minimizing methods. The study made it possible to formulate the problem of energy efficient scheduling for a set of CCs and propose an algorithm for its solution. Experiments on NPB benchmarks allowed to achieve significant reduction in energy consumption with minor increase of runtime.

**Introduction**

Running the same programs with the same input data on different computers will consume different amount of electric energy. Despite existence of various algorithmic and software methods to reduce power consumption, users prefer not to use them, since they can result a significant increase in program execution time [1]. At the same time it is shown in [2, 3] that for some classes of parallel programs, particularly with large amount of data exchange or disk operations, energy consumption can be reduced without increasing execution time.

Typically, user access to supercomputer resources is implemented using job management systems (JMS) like Slurm, PBS, LSF, SUPPZ and others. One of the key elements of JMS is a scheduler, which is maintaining the queue of user jobs and setting the order of supercomputer resources allocation. Currently several scheduling algorithms are used, the most common are «First In, First Out», Backfilling and their variations [4]. To solve the problem of reducing energy consumption of computing systems the means of monitoring the state and control of computing nodes are added to JMS functionality [5]. This functionality can be used by scheduler to make a decision to change the order of computing resources allocation, and by the JMS to implement operations to control and minimize energy consumption.

Currently two main ways of energy consumption reduction are used in JMS. The first way is based on the complete shutdown or switch to the standby mode of idle computing nodes. The order of shutdown of computing nodes is determined by the scheduler of the batch system. For example, Slurm implements a power-saving technique, when nodes which are idle for a specified period of time can be switched to power-saving mode or turned off [6]. The nodes will be returned to normal operation mode just before allocation to the parallel program. The negative effect of reducing power consumption is the increased job wait time in the queue in proportion to the load time of computational nodes.

The second way is based on the compulsory limitation of the computing nodes power when the threshold temperature or power consumption is reached [7]. For example, it is possible in Linux to adjust the frequency and voltage of the processor [8] using the DWFS mechanism embedded into the microprocessor and system utilities (like *cpufreq* or similar) inserted into the OS kernel. The disadvantage of this method of reducing power consumption is decrease in the performance of computing resources of the SCC, and, consequently, an increase in the execution time of applications.

An alternative method of energy efficient scheduling of computing resources is used in the EAS extension [9] extension of the Linux task scheduler to optimize computations on the cores of systems on big.LITTLE and DynamIQ Arm platforms. The scheduler calculates the energy consumption of each processor for executing instructions and selects the most energy efficient solution without decreasing performance. In fact, EAS allows the Linux task scheduler to take advantage of the energy efficiency of "small" processor cores in cases where this does not lead to a decrease in computing performance compared to the energy-intensive "large" cores. Despite its effectiveness, this method has not been implemented for cluster computing systems.

**Problem of energy efficient scheduling of computing resources in clusters**

Depending on the execution model, a parallel program can be run both on one or on several computing nodes (CN) of a supercomputer. Regardless of the used algorithm, the following phases of parallel program execution can be distinguished: the phase of computation, the phase of external memory access, the phase of communication between the CNs [10]. The execution time of each phase and the amount of consumed energy depends on the characteristics of the computing system.

Let us denote:

$T$ – execution time of parallel program;

$N$ – number of CNs allocated to the job;

$E_{calc,i}^{j}(t)$ - amount of consumed energy by device $i$ of the node CN$j$ at time point $t$, where $t \in [1; T], j \in [1; N]$;

$E_{CALC,\Sigma}^{j}(t)$ - amount of energy consumed by all computing devices of CN$j$ at time point $t$:

$$E_{CALC,\Sigma}^{j}(t) = \sum_i E_{calc,i}^{j}(t), \text{где } i = \{CPU, GPU, \dots\}.$$

$E_{disk}^{j}(t)$ – amount of energy consumed by CN$j$ to access to external memory at time point $t$.

$E_{net}^{j}(t)$ – amount of energy consumed by CN$j$ to send/receive data through the communication network at time point $t$.

$W^{j}(t)$ – power consumption of CN$j$ at time point $t$ can be calculated by the formula:

$$W^{j}(t) = E_{CALC,\Sigma}^{j}(t) + E_{disk}^{j}(t) + E_{net}^{j}(t)$$

Then, taking into account the designations above, the average value of the power consumption in Watts by CN ($W$) during a parallel program execution can be calculated by the formula:

$$W = \frac{\int_0^T \sum_j W^j(t)\, dt}{T}, \text{где } j \in [1, N]$$

To estimate the impact of a parallel program on the power consumption of computing systems, we propose the notion of a power consumption profile of a parallel application. Under this term we mean the set of values $(K, C)$, where:

$K$ – coefficient of the acceptable increase in the execution time $T$, in percent (%).

$C$ – coefficient of the energy consumption CN for performing computational operations in a parallel program (Joules per operation, J/op):

$$C = \frac{W}{P}$$

*P* – number of program computational operations performed by the CN in 1 second (op/s).

The problem of energy efficient scheduling is the selection of computational resources to execute parallel program when the lowest value *C* is achieved with a given threshold value *K*.

**Algorithm**

To solve the problem of computing resources energy-efficient scheduling, an algorithm was implemented that allows you to choose the optimal computing system for launching jobs. The algorithm takes into account the energy consumption of the parallel application (*C*) and the acceptable increase in the execution time of the parallel program (*K*). The proposed algorithm is based on the ideas of expanding the EAS, adapted to use in an SCC.

In general, the algorithm can be represented by the following sequence of steps.

Step 1. Generate a list of computing systems (*Systems*) for launching a parallel program.

Step 2. For all computing systems in the *Systems* list determine values of variable *C* at the previous runs of the parallel program. If the *Systems* list contains a computing system on which the parallel program has not been run before, then set the value *C* = 0.

Step 3. For all computing systems in the *Systems* list determine values of variable *T* at the previous runs of the parallel program. If the *Systems* list contains a computing system on which the parallel program has not been run before, then set the value *T=0*.

Step 4. Select from the *Systems* list the computing system with the smallest value of C at a given threshold value K. Finish the algorithm.

Let us consider an example of the algorithm work for the case with 3 computing systems, differing in performance and power consumption. Suppose that it is planned to run 5 different parallel programs for the first time. Programs are numbered as they entered the JMS queue. The specified threshold value of program runtime increase does not exceed *K*.

At the first step of the algorithm, a list of available computing systems will be generated: $CC_1$, $CC_2$, $CC_3$ (step 1). Since no parallel programs have been run on any system earlier, the value *C* = 0 (step 2) and the value *T* = 0 (step 3) will be set for all of them. Since for all computing systems the values of *C* and *T* are equal to 0, then each parallel program will be submitted on the first released computing system (step 4). After the successful completion of each parallel program, the *C* and *T* values are stored for the selected computing system (see Table 1 and Table 2).

Table 1

|           | $CC_1$    | $CC_2$    | $CC_3$    |
|-----------|-----------|-----------|-----------|
| Program 1 | $C_{1,1}$ |           |           |
| Program 2 | $C_{2,1}$ |           |           |
| Program 3 |           | $C_{3,2}$ |           |
| Program 4 |           | $C_{4,2}$ |           |
| Program 5 |           |           | $C_{5,3}$ |

Table 2

|           | $CC_1$    | $CC_2$    | $CC_3$    |
|-----------|-----------|-----------|-----------|
| Program 1 | $T_{1,1}$ |           |           |
| Program 2 | $T_{2,1}$ |           |           |
| Program 3 |           | $T_{3,2}$ |           |
| Program 4 |           | $T_{4,2}$ |           |
| Program 5 |           |           | $T_{5,3}$ |

Table 3

|           | $CC_1$    | $CC_2$    | $CC_3$    |
|-----------|-----------|-----------|-----------|
| Program 1 | $C_{1,1}$ | $C_{1,2}$ | $C_{1,3}$ |
| Program 2 | $C_{2,1}$ | $C_{2,2}$ | $C_{2,3}$ |
| Program 3 | $C_{3,1}$ | $C_{3,2}$ | $C_{3,3}$ |
| Program 4 | $C_{4,1}$ | $C_{4,2}$ | $C_{4,3}$ |
| Program 5 | $C_{5,1}$ | $C_{5,2}$ | $C_{5,3}$ |

Table 4

|           | $CC_1$    | $CC_2$    | $CC_3$    |
|-----------|-----------|-----------|-----------|
| Program 1 | $T_{1,1}$ | $T_{1,2}$ | $T_{1,3}$ |
| Program 2 | $T_{2,1}$ | $T_{2,2}$ | $T_{2,3}$ |
| Program 3 | $T_{3,1}$ | $T_{3,2}$ | $T_{3,3}$ |
| Program 4 | $T_{4,1}$ | $T_{4,2}$ | $T_{4,3}$ |
| Program 5 | $T_{5,1}$ | $T_{5,2}$ | $T_{5,3}$ |

Computing resources will be allocated on previously unused computing systems to run programs until tables 1 and 2 are filled. This example requires no more than 3 runs of each program (see Tables 3 and 4). After that, the CC will be chosen according to the values of *C, T* and *K* corresponding to each program. For example, if for the Program 1 the least execution time is reached on $CC_3$, and the least energy consumption is reached on $CC_1$, then the program will be submitted on CC for which the following conditions are met.

$$\begin{cases} T_{1,i} \leq T_{1,3} + K \\ \phantom{T_{1,i}} C_{1,i} \geq C_{1,1} \end{cases}$$

Table 5 illustrates the operation of algorithm with all 5 previously run programs, with addition of Program 6 that was executed once and Program 7 that was never submitted on considered clusters.

Table 5

|  | Energy consumption *C*, J/op | | | Runtime, sec. | | | Acceptable increase | Allocated cluster |
|---|---|---|---|---|---|---|---|---|
|  | $CC_1$ | $CC_2$ | $CC_3$ | $CC_1$ | $CC_2$ | $CC_3$ | K |  |
| Program 1 | 0,0015 | 0,002 | 0,001 | 550 | 500 | *700* | 10% | $CC_1$ |
| Program 2 | 0,0012 | 0,0015 | 0,0013 | 500 | 350 | *650* | 30% | $CC_2$ |
| Program 3 | 0,0013 | 0,0019 | 0,0011 | 700 | 500 | 900 | 90% | $CC_3$ |
| Program 4 | 0,0055 | 0,0075 | 0,006 | *180* | 100 | 120 | 50% | $CC_3$ |
| Program 5 | 0,005 | 0,0055 | 0,0045 | *5000* | 4500 | *6000* | 0% | $CC_2$ |
| Program 6 | 0 | 0 | 0,005 | 0 | 0 | 150 | 15% | $CC_1$ |
| Program 7 | 0 | 0 | 0 | 0 | 0 | 0 | 25% | $CC_3$ |

The calculation of the energy consumption value and the determination of the parallel program execution time are performed automatically in accordance with the method presented in [11].

**Implementation**

The presented algorithm was implemented in the C ++ programming language and integrated into the command interface for job submission of the SUPPZ job management systems [12]. A modified version of the *mpirun* command is used in the SUPPZ to run parallel programs. The user specifies in parameters of this command number of processors required by the job, maximum occupancy time of computing resources of the system, the type of computing resources used, and name of the executable file with input arguments.

When the *mpirun* command is invoked, the hash value of the executable file is calculated. The hash value is stored in a separate file and used as a unique identifier for the program. The hash value, as well as the *mpirun* arguments used at the program submission, are saved to the database. Next, the algorithm for selecting computing systems is initiated.

If, at parallel program submission the user specified the type of computing resource to be used, then the result of the algorithm operation will just a notification. A computing system with the lowest power consumption will be offered to the user. If the user did not specify the type of computing resource, then the parallel program will be automatically queued on the computing system with the lowest power consumption for the given application. The parameter *K* value, which indicated the allowable excess of the parallel program runtime, can be specified by the administrator or calculated automatically before starting the algorithm. So, if a parallel program was executed before and its runtime (*T*) did not exceed the ordered time of computing resources ($T_{max}$), then the value of *K* is calculated by the formula:

$$K = \frac{T_{max}}{T}$$

**Experiments**

To check the practicability of the developed algorithm as part of SUPPZ, the authors carried out a series of experiments on NAS Parallel Benchmarks 3.3 (NPB) BT, EP, IS, LU, and SP tests, class D. This class was chosen due to the sufficient load of computational nodes, performance and the use of computational resources in during programs execution.

Experimental platform consists of supercomputers MVS-10P MP2 KNL (KNL), MVS-10P OP BRD (Broadwell), MVS-10P OP SKX (Skylake) and MVS-10P OP CLK (Cascade Lake) at the JSCC RAS [13].

The listed supercomputing systems have different performance and power consumption. Table 6 shows the number of cores and the number of CNs allocated by SUPPZ to run each test program.

Table 6

NPB run parameters

| Benchmark | Number of cores | Number of allocated CNs | | | |
|---|---|---|---|---|---|
| | | Broadwell | Cascade Lake | KNL | Skylake |
| BT | 144 | 5 | 3 | 2 | 4 |
| EP | 144 | 5 | 3 | 2 | 4 |
| IS | 256 | 8 | 6 | 4 | 8 |
| LU | 256 | 8 | 6 | 4 | 8 |
| SP | 256 | 8 | 6 | 4 | 8 |

The efficiency of the algorithm was evaluated basing of power consumption and the runtime of the entire test set. Figures 1 and 2 show the result of comparing power consumption (W) and execution time (T) for a set of simultaneously launched tests from Table 1 for different modes of the algorithm work. For example, Alg (0) corresponds to the mode with the value of the parameter *K* equal to 0, and Alg (85) corresponds to the mode of operation with the value of the parameter *K* equal to 85.

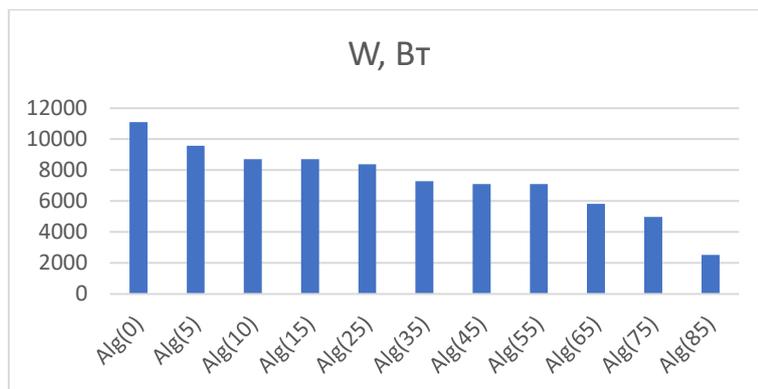

Fig. 1. Dependence of energy consumption on algorithm parameters

The presented results show that even with a slight increase in the parameter *K* value (from 5 to 10%), a significant reduction in energy consumption is achieved. For the test suite shown in Table 1, it was possible to reduce power consumption by an average of 21.5%, while the test suite execution time increased by 3.8%. Note that for all tests except LU, it was possible to achieve a reduction in power consumption with an allowable increase in program execution time by less than 5%.

In Figures 3 and 4 you can see the results of comparing energy consumption and execution time for each NPB benchmark from the given set, depending on the value of the parameter *K*.

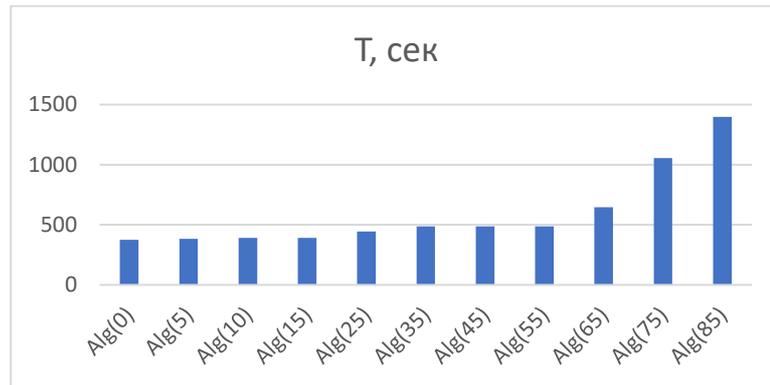

Fig. 2. Dependence of NPB tests set runtime on algorithm parameters

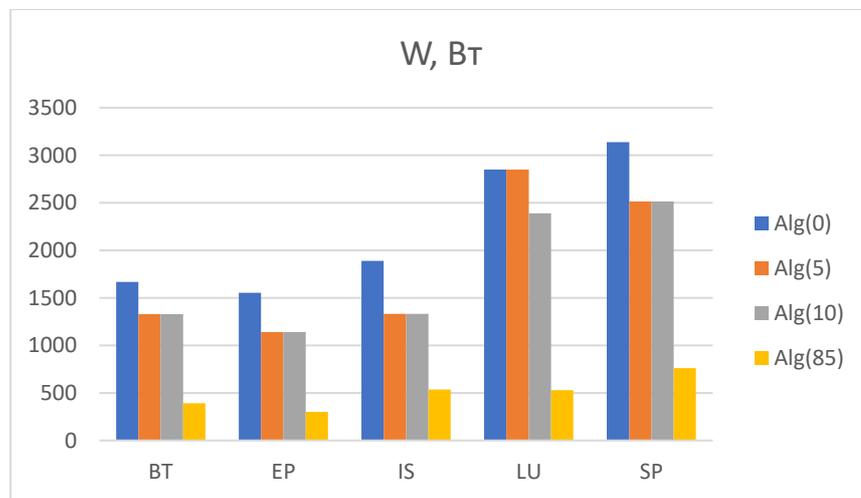

Fig. 3. Dependence of NPB tests set energy consumption on algorithm parameters

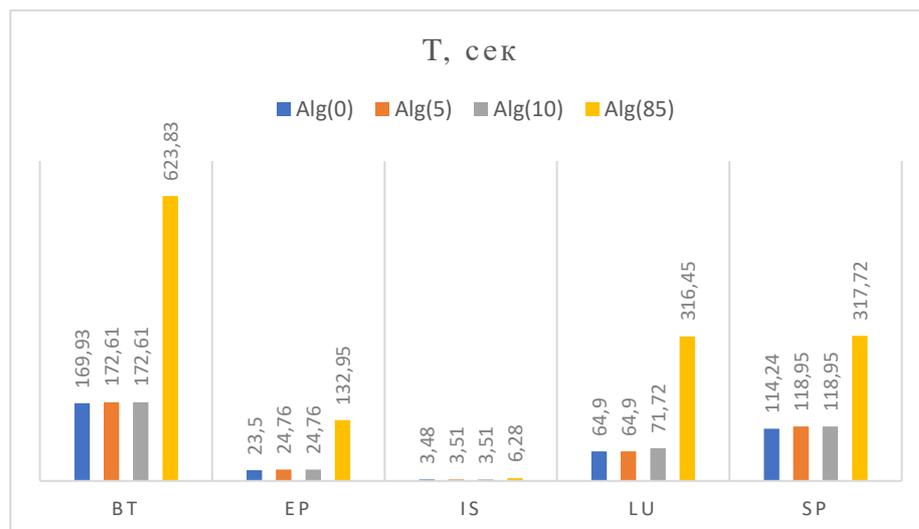

Fig. 4. Dependence of NPB benchmarks runtime on algorithm parameters

**Conclusion**

The results of the study lead us to the conclusion that the energy efficient scheduling algorithm is useful for batch systems like SUPPZ. Although efficiency of the presented algorithm vary for different benchmarks, for a set of parallel programs submitted together we achieved substantial reduction in energy consumption with minor increase in runtime (less than 10%).

Since the results presented in the paper were obtained on test programs, the authors plan to continue the discussed research using actual parallel applications of SCC users. We also plan to estimate the change in the runtime of parallel programs, taking into account the wait time in the queue for a set of computing resources at SCC.